\documentclass[prd,twocolumn]{revtex4}
\usepackage{graphicx}
\usepackage{dcolumn}
\usepackage{bm}
\begin{document}

\title{A numerical examination of an evolving black string horizon}

\author{David Garfinkle}
\email{garfinkl@oakland.edu}
\affiliation{Department of Physics, Oakland University, Rochester, MI 48309}

\author{Luis Lehner}
\email{lehner@baton.phys.lsu.edu}
\affiliation{Department of Physics and Astronomy, Louisiana State 
University, Baton Rouge, LA 70810}

\author{Frans Pretorius}
\email{fransp@tapir.caltech.edu}
\affiliation{Theoretical Astrophysics, California Institute of 
Technology, Pasadena, CA 91125}

\begin{abstract}
We use the numerical solution describing the evolution of a perturbed
black string presented in \cite{Simulation} to elucidate the intrinsic 
behavior of the  horizon.  It is found that by the end of the
simulation, the affine parameter on the horizon has become very large and 
the expansion and shear of the horizon in turn very small. This 
suggests the possibility that the horizon might pinch off in infinite 
affine parameter.
 
\end{abstract}
\maketitle

\section{Introduction}

While in four-dimensional spacetimes black holes are stable, it has been shown by Gregory and
Laflamme\cite{GL} that at least some of their higher dimensional analogs, in particular black strings,
are {\it linearly} unstable.  As opposed to the situation in four-dimensions, at the linear level
perturbations of these black strings grow exponentially above some critical length
of the string. Although this picture has been well understood through a variety
of analysis \cite{GL,GL2,gubser,sorkin,reall,sarbachlehner}, it
is not yet known what an unstable black string 
evolves to.  In \cite{GL}, using the linear instability of the string coupled to
entropy considerations, it was argued that an unstable black string
would ``pinch off'' and form a set of black holes which would later merge.
This scenario, at the classical level, must give rise to a violation of
cosmic censorship.  However, this possibility was made doubtful by more
recent work of Horowitz and Maeda (HM)~\cite{Gary}. There, a theorem is proved using the 
properties of expansion, shear and affine parameter of the geodesic
generators of the black string horizon.  This theorem asserts that a 
pinch off in finite affine parameter could only occur if a 
singularity formed outside the horizon.
Under the assumption that this, perhaps 
too drastic scenario, is unlikely, the only other option left open for the event horizon pinching-off is
that it does so in {\it infinite affine parameter}.  This possibility, although
mathematically possible, was argued against by HM. Their argument is based
on one main observation together with a plausibility argument. 
The observation is that although their
theorem does indeed allow for a pinch-off in infinite time, the rate,
with respect to affine parameter, at which circles at $r=const$ can
decrease in size is extremely slow. HM then argue that 
if the system decays so slowly, its dynamics
must go through a series of essentially static non-uniform strings, and it then might just stay at one
of these solutions. 
Therefore, instead of the pinch-off scenario, HM conjectured
that the black string would evolve into a black string that is 
static but not uniform in the $z$ direction: a static wiggly black
string.  Families of  static wiggly black strings were found numerically by
Wiseman~\cite{Toby}, however, all the solutions of \cite{Toby} had too 
large a mass to be the endpoint of the evolution of the unstable black
string as conjectured in \cite{kol}. Note however the recent work by Sorkin \cite{sorkin} (and related
discussions in \cite{sorkinkol}) where it is discussed how for large enough
spherical dimensions the static solutions found do indeed have lower mass,
and could in principle be those conjectured by \cite{Gary}.

In an attempt to resolve the issue of the fate of the linearly unstable black string, in \cite{Simulation}
a numerical discretization of the Einstein equations in five-dimensions was presented.
An SO(3) symmetry was assumed,
effectively making the problem 2+1 dimensional, and hence one that could be solved
on contemporary computer systems. This code was employed
to study the evolution of slight perturbations of a static, uniform black string.
The simulation tracked the behavior of the system well into the non-linear regime,
though unfortunately it
could not be extended far enough to elucidate the final fate of
the string. Nevertheless, the numerical solution revealed significant and
rich dynamics, illustrating, in particular, that the string evolves toward
a shape resembling large spherical black holes connected by thin strings.
In terms of areal radius, the ratio of the maximum to minimum
radius was about $10$ by the end of the simulation.

In this paper, we re-examine the data of \cite{Simulation}, paying closer attention
to the analysis of \cite{Gary} and examine relevant features of the solution. In particular,
we numerically find the expansion, shear and affine parameter of 
the black string horizon and discuss their behavior in connection to the
conjecture of \cite{Gary} where a scenario is ruled out as improbable, though as we
argue here, it might very well be possible.  
The methods used are presented in section
II, results in section III and conclusions in section IV.

\section{Methods}
Our starting point is to consider a spacetime described by
the metric provided by the numerical simulation of \cite{Simulation}. For
simplicity we adopt the same gauge conditions used there (to avoid performing
a numerical coordinate transformation, which in general would involve interpolating functions
in space and time).  The metric used in \cite{Simulation} takes the form
\begin{eqnarray}
d {s^2} &=& (- {\alpha ^2} + {\gamma _{AB}}{\beta ^A}{\beta ^B}) d{t^2}
+ 2 {\gamma _{AB}}{\beta ^A} d {x^B} d t \nonumber \\
& &+ {\gamma _{AB}}d{x^A}d{x^B} 
+ {\gamma _\Omega} d {\Omega ^2}
\end{eqnarray}
where ${x^A}=(r,z)$ and $d{\Omega ^2}$ is the unit two-sphere metric.
Our goal is to extract intrinsic properties of the event horizon and its
generators. The horizon can be described by a surface $r=R(t,z)$. 
To extract the sought-after information, the first step is  to find the geodesic
generators of the horizon.  To this end, we construct the vector $k_a$  
normal to the horizon defined as $k_a = ({k_t},{k_r},{k_z})=(-{\dot R},1,-{R '})$.
Here an overdot denotes differentiation with respect to $t$ and a prime
denotes differentiation with respect to $z$. Since $k^a$ is null, it is tangent to the 
generators of the horizon; these are given by
\begin{equation}
{{dz}\over {dt}} = {\alpha ^2} \left ( {{{\gamma ^{rz}} - {\gamma ^{zz}}
{R '}}\over {{\beta ^r} + {\dot R}}} \right )
\end{equation}
Note however that these geodesics are not affinely parameterized. To obtain
the affinely parameterized geodesics one can exploit the fact that 
their tangent vectors ${l^a}$ obey the simple relation ${l^a}= {e^{-\nu }} {k^a}$
where $\nu $ is defined by
\begin{equation}
{{d\nu}\over {dt}} = {{{\alpha ^2} h} \over {{\beta ^r} + {\dot R}}}
\end{equation}
and $h$ is given by
\begin{equation}
h = {1 \over 2} {\partial \over {\partial r}} \left ( 
{\gamma ^{rr}} + {\gamma ^{zz}} {{({R'})}^2} - 2
{\gamma ^{rz}}{R'} - {\alpha ^{-2}} {{({\dot R}+{\beta ^r})}^2}
\right )
\end{equation}  
Thus, the affine parameter $\lambda$ of these geodesics is found by
integrating
\begin{equation}
{{d\lambda }\over {dt}} = {{{\alpha ^2} {e^\nu}}\over 
{{\beta ^r} + {\dot R}}} \, .
\end{equation}
Next we turn our attention to obtaining the expansion and shear of
these affinely parametrized generators. Since the spacetime is spherically
symmetric, these quantities are given by just two components 
of ${\nabla _a}{l_b}$: Let $x^a$ be a unit vector
in the two-sphere direction and let $y^a$ be a unit vector orthogonal to
$k^a$ and to the two-sphere directions.  Define the quantities $A$ and
$B$ by
\begin{eqnarray}
A = 2 {x^a}{x^b} {\nabla _a}{l_b} 
\\
B = - {y^a}{y^b} {\nabla _a}{l_b}
\end{eqnarray}
which measure the distortion along spherical and longitudinal directions
respectively. Then, 
the expansion and squared shear are given in terms of $A$ and $B$ by
\begin{eqnarray}
\theta &=&  A - B \\
{\sigma ^{ab}}{\sigma _{ab}} &=& {1\over 6} {{(A+2B)}^2} \, .
\end{eqnarray}
The equations above determine the affinely parametrized generators of
the horizon and its expansion and shear. As we will see later, due to the
exponential behavior of these quantities, it will
turn out to be convenient to consider the logarithm of the affine parameter
$s \equiv \ln \lambda $ and rescaled quantities defined as follows
\begin{equation}
({\tilde \theta },{{\tilde \sigma }_{ab}}, {\tilde A}, {\tilde B})
\equiv \lambda (\theta , {\sigma _{ab}}, A, B)  
\end{equation}
It then follows that 
\begin{eqnarray}
{\tilde \theta } &= & {\tilde A} - {\tilde B}
\label{thetatilde}
\\
{{\tilde \sigma}^{ab}}{{\tilde \sigma}_{ab}} &=& {1 \over 6}
{{({\tilde A} + 2 {\tilde B})}^2}
\label{sigmatilde}
\\
{{d s} \over {d t}} &=& {{{\alpha ^2} {e^{(\nu - s)}}}\over 
{{\beta ^r} + {\dot R}}}  
\end{eqnarray}
The quantity $\tilde A$ is found by
\begin{equation}
{\tilde A} =  {e^{(s-\nu )}} {k^a} {\nabla _a}
(\ln {\gamma _\Omega})
\end{equation}
To find $\tilde B$ we note that there is freedom to add a multiple
of $k^a$ to $y^a$ and use that freedom to make $y^t$ vanish.  It
then follows that the components of $y^a$ are 
\begin{equation}
{y^z} = {{\left [ {\gamma _{zz}}+2{\gamma _{rz}}{R'} + {\gamma _{rr}}
{{({R'})}^2}\right ] }^{-1/2}}
\end{equation}
and ${y^r}= {y^z} {R '}$, from which 
\begin{equation}
{\tilde B} =  {e^{(s - \nu )}} \left ( {{({y^z})}^2} {R''} + {k_b}
{\Gamma ^b _{ac}}{y^a}{y^c}\right )
\end{equation}

Finally, we point out the following observation which will be crucial
in the numerical evaluations to be carried out later.
As we have seen above,  $\theta$ and ${\sigma}_{ab}$ (and their scaled counterparts) 
can be calculated  independently. However they are related by  Raychaudhuri's equation, which
for the five-dimensional vacuum null case is
\begin{equation}
{{d\theta}\over {d\lambda}}+ {\textstyle {1 \over 3}} {\theta ^2}
+{\sigma ^{ab}}{\sigma_{ab}} = 0 \, ,
\label{Ray}
\end{equation}
while the corresponding equation for the rescaled quantities is
\begin{equation}
{{d {\tilde \theta }}\over {d s}} - {\tilde \theta } + 
{\textstyle {1\over 3}} {{\tilde \theta }^2} + {{\tilde \sigma}^{ab}}
{{\tilde \sigma}_{ab}} = 0.
\label{Rayscale}
\end{equation}

\section{Results}
The simulations performed in \cite{Simulation} provide a (partial)
description of the spacetime describing a slightly
perturbed black string that evolves to a state where the black
string is quite deformed as judged by the apparent horizon. 
The surface that we use here as the event horizon is the boundary
of the causal past of the region exterior to the apparent horizon at the latest time in
the simulation (calculated using a generalization of the technique described
in \cite{Simulation}), and it turns out that this surface is very well
approximated by the apparent horizon for the entire length of the simulation.
We initialize the affine parameter to $1$ (hence $s=0$) at the beginning of the
simulation.
Among all generators, the one of particular interest
is the one corresponding to the minimum
radius, since it is there where any pinch off would occur.  
Due to the reflection symmetry of the initial data, 
the set of points on the horizon that at each time correspond
to the minimal radius form a geodesic generator, and that
generator stays at constant $z$.  

\begin{figure}
\includegraphics[scale=0.4]{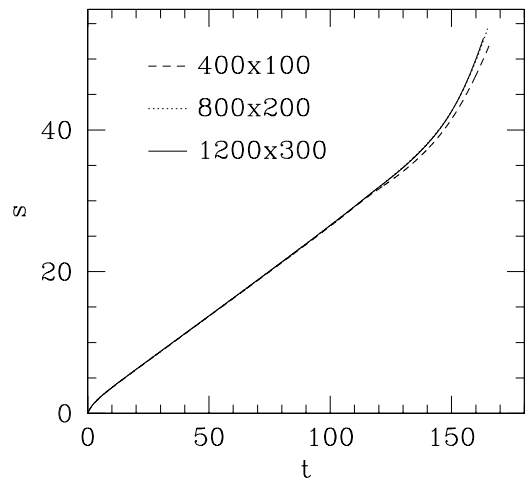}
\caption{\label{fig1} $s$ vs $t$ for the horizon generator of minimum radius, 
from three simulations of identical initial
data though differing resolution (the curves are labelled by $N_r$x$N_z$,
where $N_r$($N_z$) is the number of grid points in the $r$($z$) direction).
We show results from three different resolutions to demonstrate that
we are in the convergent regime. Recall that the affine parameter $\lambda=e^s$,
and therefore $\lambda$ is growing very rapidly with simulation time $t$.}
\end{figure}

Figure \ref{fig1} shows plots of $s$ as a function of the 
simulation time coordinate $t$ for the generator of minimum
radius, from simulations with three different resolutions (to give
some measure of the error in the results).
Note that $s$ has reached $50$ near the end of the simulation.
This corresponds to an affine parameter of about ${10}^{21}$.  Thus,
though the simulation has only taken a moderate amount of simulation
time, it has gone for an enormous amount of affine parameter. 

\begin{figure}
\includegraphics[scale=0.42]{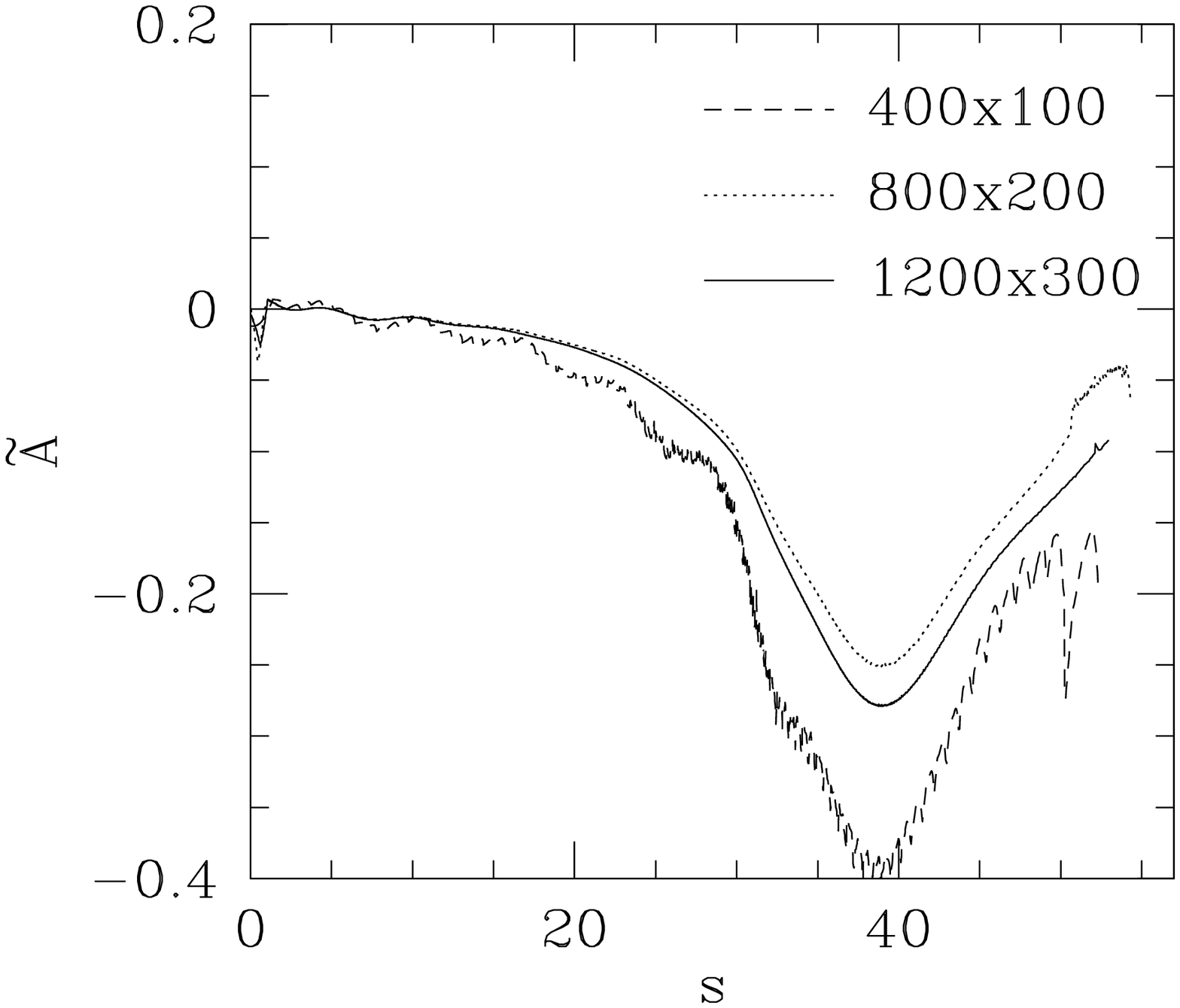}
\includegraphics[scale=0.42]{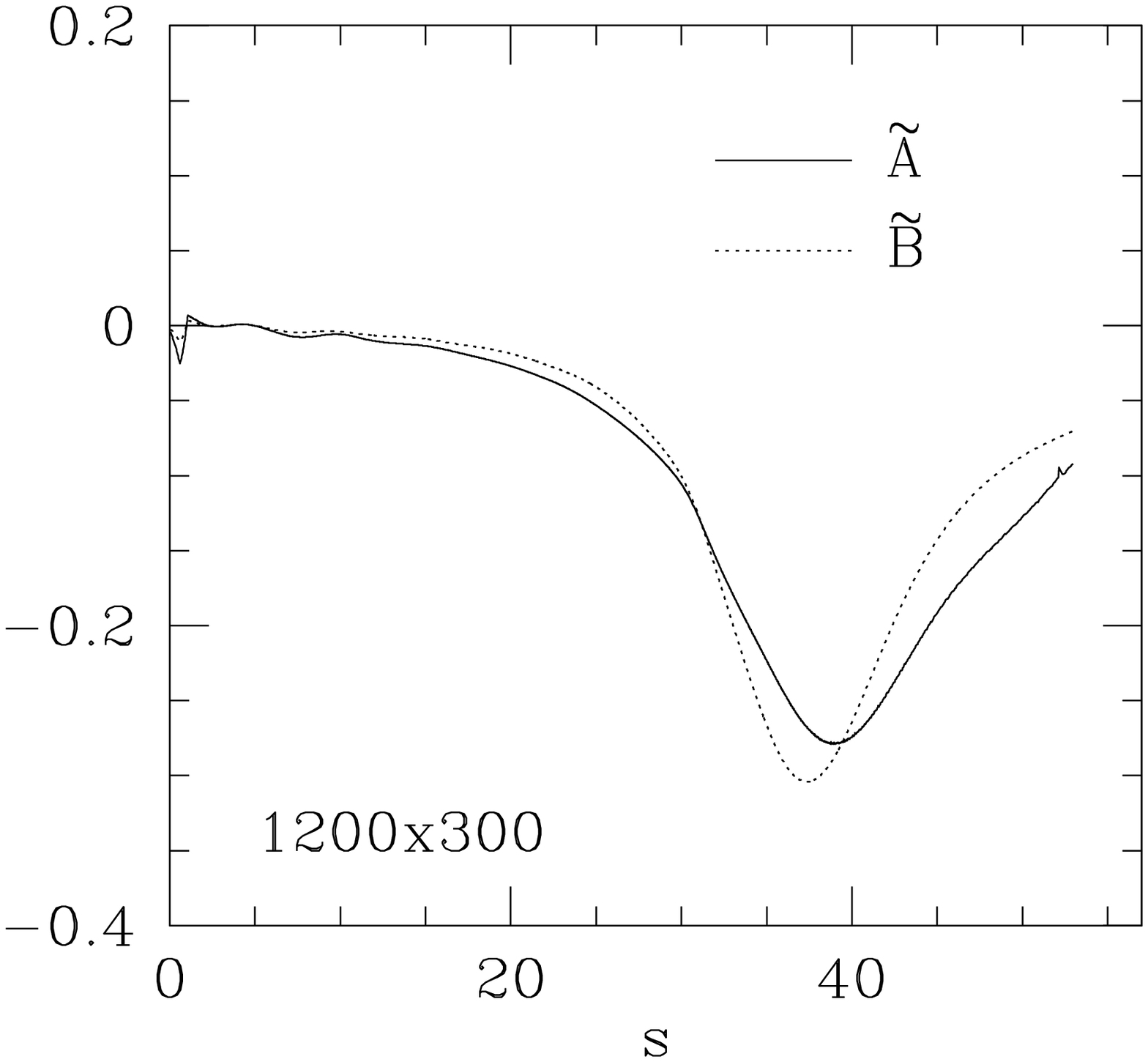}
\caption{\label{fig2} The plot at the top  depicts $\tilde A$ as a function
of $s$, while the plot at the bottom
shows $\tilde A$ and $\tilde B$ vs. $s$ from the highest resolution 
simulation, both for the horizon generator of minimum radius. 
What the latter plot demonstrates is that $\tilde A$ and $\tilde B$
are very similar in magnitude; in particular the difference $\tilde A - \tilde B$
(which is just $\tilde \theta$) at late times is completely
dominated by numerical error, the magnitude of which can be estimated by using the
data from the three simulations.
Specifically, the error in $\tilde A$ from
the highest resolution simulation is quite small initially, is near $5\%$ near the minimum of
$\tilde A$, and has grown to around $40\%$ by the end of the simulation ($\tilde B$ has similar error).}
\end{figure}

Figure \ref{fig2} shows plots of ${\tilde A}$ and ${\tilde B}$
vs $s$ for the generator of minimum radius.
Note that due to the rescaling this means that by the end of the simulation,
the unrescaled expansion $\theta$ and shear $\sigma _{ab}$ are of
order ${10}^{-22}$.  Note too, that though the behavior of ${\tilde A}$
and ${\tilde B}$ as a function of $s$ is quite dynamic throughout the
simulation, since $d/d\lambda = {\lambda ^{-1}} d/ds$ it follows that 
viewed as functions of $\lambda$ even the rescaled expansion and shear
would be regarded as slowly varying.

\begin{figure}
\includegraphics[scale=0.4]{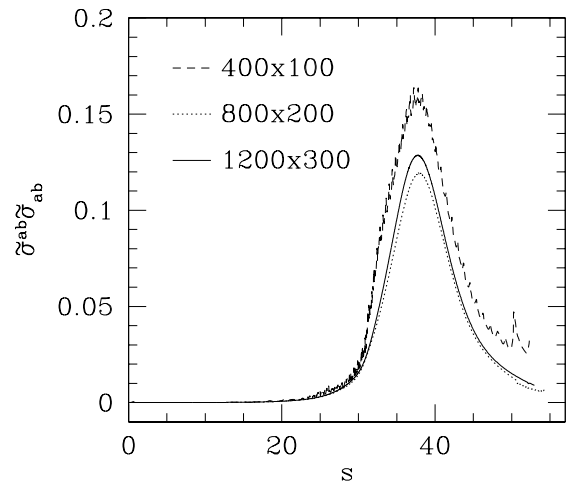}
\caption{\label{fig3} ${{\tilde \sigma}^{ab}}{{\tilde \sigma}_{ab}}$ 
vs. $s$ for the horizon generator of minimum radius.}
\end{figure}

We now turn to the behavior of the rescaled quantities ${\tilde \theta}$
and ${\tilde \sigma}_{ab}$. We can calculate the shear by simply evaluating
(\ref{sigmatilde}) along the generator of minimum radius: Figure \ref{fig3} shows
the result of this as a function of $s$ for the
three simulations. In principle we can also calculate ${\tilde \theta}$ in a 
similar manner using (\ref{thetatilde}), which says ${\tilde \theta}$ is
the {\em difference} between ${\tilde A}$ and ${\tilde B}$. However, the problem
with this is that ${\tilde A}$ and ${\tilde B}$ are very close to one another
in magnitude, as can be seen from Figure \ref{fig2}, 
and so the difference
will be dominated by numerical error. In fact, from the plot of ${\tilde A}$ in 
Figure \ref{fig2}, and the assumed second order convergence of the numerical
scheme, one can estimate that the error in ${\tilde A}$ (and similarly for ${\tilde B}$),
is on the order of $5\%$ at intermediate times (near the minimum of ${\tilde A}$)
and grows to as much as $40\%$ at late times. Therefore at late times the
difference ${\tilde A}-{\tilde B}$ will be completely swamped by numerical
error. 

%
%
%

In order to alleviate this problem, we use Raychaudhuri's equation 
to integrate $\tilde \theta$. However, in the form of (\ref{Rayscale})
there is in general no stable direction of integration due to the linear and
quadratic terms in $\tilde \theta$; in other words, numerical errors in $\tilde \theta$ could cause a blow-up
integrating in either direction in $s$. Therefore, we eliminate
the $\tilde \theta^2$ term using (\ref{thetatilde}), and ${{\tilde \sigma}^{ab}}{{\tilde \sigma}_{ab}}$
using (\ref{sigmatilde}), though we keep the
linear term in $\tilde \theta$ to give

\begin{equation}
{{d {\tilde \theta }}\over {d s}} - {\tilde \theta } + 
\frac{1}{2}\tilde A^2 + \tilde B^2 = 0.
\label{Rayscale_b}
\end{equation}

\begin{figure}
\includegraphics[scale=0.4]{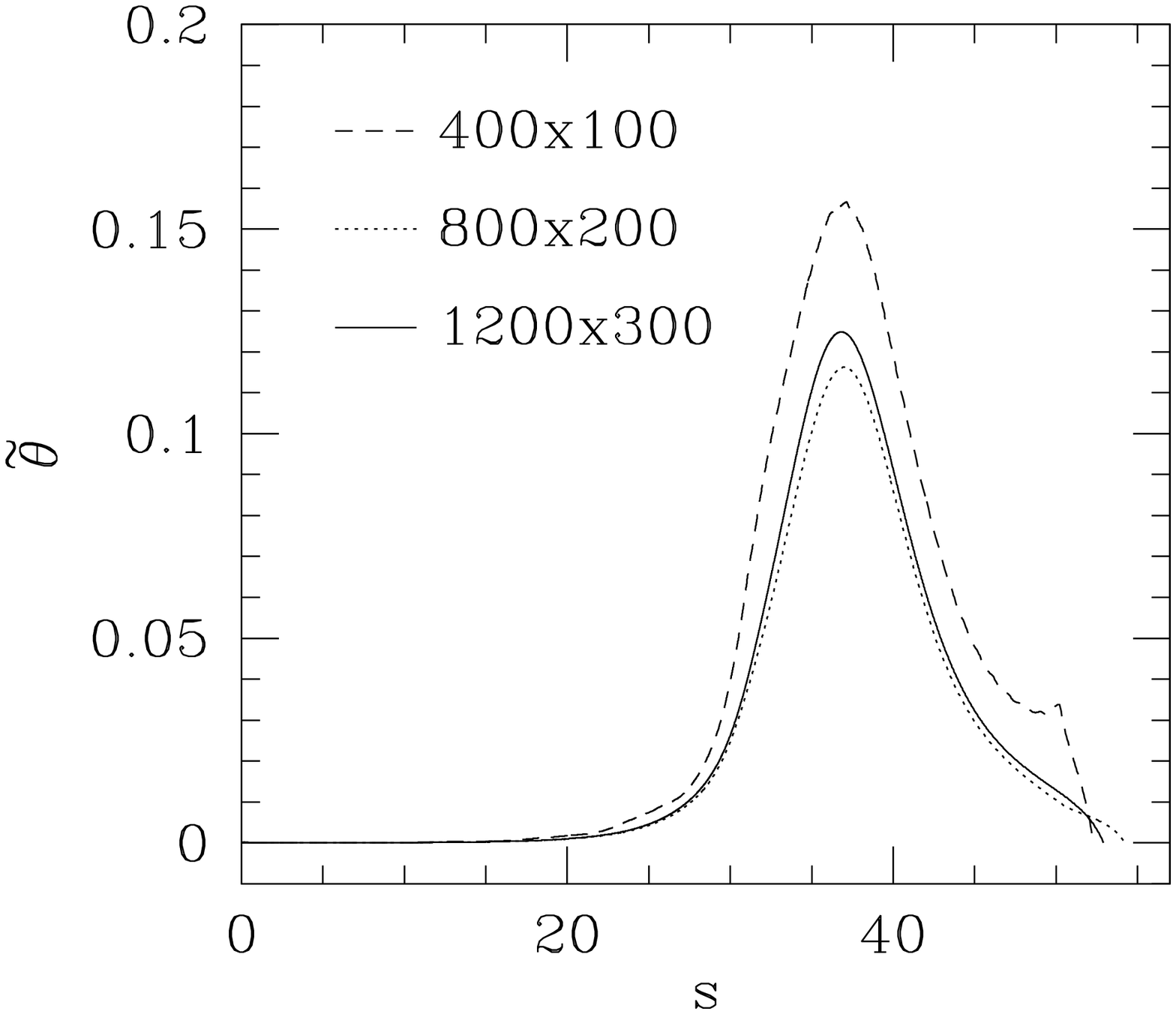}
\includegraphics[scale=0.4]{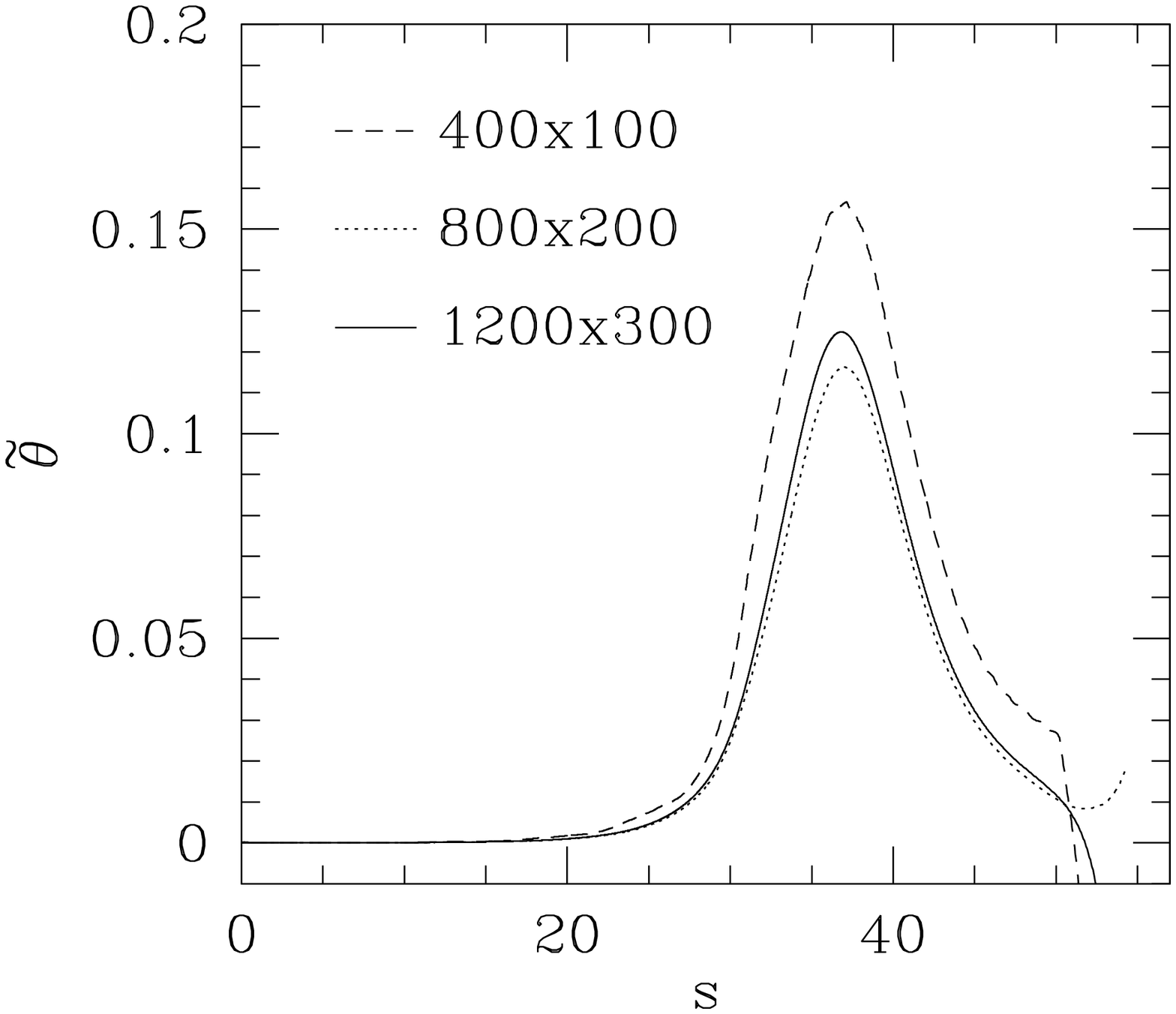}
\caption{\label{fig4} $\tilde \theta$ vs. $s$ for the horizon generator of minimum radius, calculated by integrating
(\ref{Rayscale_b}). For the integrations shown on the top figure,
the initial value of $\tilde \theta$ (which is at the {\em largest} $s$, as we integrate from large to
small $s$) is $0$, while on the bottom figure the initial $\tilde \theta$
was calculated using (\ref{thetatilde}). We show curves calculated using two different
sets of initial conditions to demonstrate that the integration is fairly insensitive
to the initial value of $\tilde \theta$; hence, expect near the end of the
simulation, the error in the integrated $\tilde \theta$ is comparable to that
of ${{\tilde \sigma}^{ab}}{{\tilde \sigma}_{ab}}$ shown in Figure \ref{fig3}---specifically
around $10\%$ near the maxima for the highest resolution simulation, assuming
second order convergence.}
\end{figure}

In this form there is a stable direction of integration from large to small $s$, and the
only `sources' in $\tilde A$ and $\tilde B$ are quadratic and of the same sign, hence 
we are not as susceptible to numerical errors arising from the difference of similar numbers.
One potential difficulty is in specifying initial conditions for the integration,
which we need to do at the end of the simulation where $\tilde \theta$ obtained from (\ref{thetatilde}) has the 
largest error. Fortunately the integration seems to be fairly insensitive
to the initial condition in $\tilde \theta$, as illustrated in Figure \ref{fig4}, where
we show the results of integration of (\ref{Rayscale_b}) starting from both 
$\tilde \theta$ calculated using (\ref{thetatilde}), and $\tilde \theta=0$ (which is within the
numerical margin of error at late times).

\section{Conclusions}

We have found the behavior of the horizon of a black string in terms of
the affine parameter, expansion and shear of the generators.  
The most significant aspects of our results are the extremely large values
of the affine parameter, and correspondingly the extremely small values
of the (unrescaled) expansion and shear. In hindsight, this
result is not too surprising as it is well known that for a 
Schwarzschild black hole there is an exponential relation between
the Killing time coordinate and the affine parameter of the horizon
generators.
Since the straight black string is just the Schwarzschild
spacetime crossed with a circle, and since the spacetime treated here
is initially just a small perturbation of the straight string, it appears
natural that very large values of the affine parameter are reached.
The immediate conclusion is that in some sense the logarithm of the
affine parameter is a more natural ``dynamical time'' than the affine
parameter.

What do our results say about the final fate of the unstable black
string?  They certainly do not yet resolve the issue.  
The same three possible
alternatives are still there: the string can (i) evolve to a 
static wiggly string, (ii) pinch off in finite affine parameter
or (iii) pinch off in infinite affine parameter.  It is clear both
from \cite{Simulation} and from our work that the simulation has not
gone far enough to read off the asymptotic behavior of the string.
For that, a more robust simulation is needed.  However, our
work does shed extra light on the status of the alternatives, 
in particular that of alternative (iii).  
Recall that in \cite{Gary} it was correctly pointed out
that alternative (iii) requires very small (unrescaled) expansion and
shear and very slow dependence on affine parameter.  Our results here
show that precisely this is the sort of behavior that occurs, at
least in some regime of the evolution of the black string. Moreover,
this behavior is still in a regime of non-trivial dynamics.

In fact, one might argue that alternative (iii) is the most plausible
for lower spherical dimensions of the spacetime {\it if the
only non-uniform static solutions are those found in \cite{Toby}}.
This fact would make alternative (i) unlikely as these solutions
are too massive for spacetimes with dimension lower than 13. 
Additionally, option (ii) requires a singularity
forming outside the horizon, and no hints of such behavior developing
were seen in the simulations of \cite{Simulation}.
Note that under alternative (iii) one might still have different scenarios.
For instance, the `minimal' circle might continue to shrink slowly with
the trend observed at the late stages of the simulations of \cite{Simulation}.
This requires the neighborhood of the string not to be describable as a linearly perturbed
uniform string since the ratio of length/area is above the critical length.
On the other hand, if this is indeed the case, then the string may pinch off 
through a cascading sequence of instabilities on ever smaller scales, which
again could happen in finite asymptotic time though {\it infinite affine time}.

In a more quantitative vein, we can consider the possible asymptotic
behavior of the (rescaled) expansion and shear.  Let $p$ and $c$
be positive constants.  Then it is consistent with equation 
(\ref{Rayscale}) for both $\tilde \theta$ and 
${{\tilde \sigma}^{ab}}{{\tilde \sigma}_{ab}}$ to asymptotically
approach ${c^2} {s^{-2p}}$.  This would result
from both $\tilde A$ and $\tilde B$ approaching $ - c {s^{-p}}$
asymptotically, but their difference approaching 
${c^2} {s^{-2p}}$.  That this sort of behavior is possible was
recognized in \cite{Gary} but there it was thought that such 
behavior was somehow unnaturally slow and indicative of
an approximately static solution.  If $ p \le 1$ then this
behavior results in pinch off in infinite affine parameter.  
If in addition, $p > 1/2$ then this pinch off occurs with  
finite $\int \theta d \lambda$ as assumed in \cite{Gary}.

If such a power law is the asymptotic behavior of the unstable
black string, then the simulations have not yet run for long enough to be 
sufficiently far into the asymptotic regime to try to calculate the exponents.
Nonetheless, in Figures \ref{fig3} and 
\ref{fig4} the quantities ${{\tilde \sigma}^{ab}}{{\tilde \sigma}_{ab}}$
and $\tilde \theta$ peak and then decrease.  This decrease may be the
beginning of a power law tail to be found in an improved black
string simulation.

\section{Acknowledgements}

We would like to thank Rob Myers, Eric Poisson,
Gary Horowitz, Toby Wiseman, Donald Marolf and Jorge Pullin
for helpful discussions.
We are especially grateful to Matt Choptuik,
Inaki Olabarrieta, Roman Petryk and Hugo Villegas who,
together with LL and FP, were involved in the project\cite{Simulation} that
produced the data used here.
Also, the authors would like 
to thank the Perimeter Institute for hospitality and to acknowledge 
discussions with the participants of the ``New Horizons'' workshop at the Perimeter
Institute (April 2004) during which this project began.
Research at the Perimeter Institute is supported in part
by funds from NSERC of Canada and the Ontario Ministry of
Economic Development and Trade. Additionally, DG and FP 
would like to thank the Horace Hearne Laboratory for 
Theoretical Physics for hospitality. 

The original simulations that produced the numerical solutions
analyzed in this paper were performed on
(i) the {\tt vn.physics.ubc.ca} cluster which was funded by the Canadian Foundation for
Innovation (CFI) and the BC Knowledge Development Fund; (ii) {\tt LosLobos} at
Albuquerque High Performance Computing Center
(iii) The high-performance computing facilities within LSU's Center for Computation and Technology, which is funded through
Louisiana legislative appropriations, and (iv) The {\tt MACI} cluster
at the University of Calgary, which is funded by the Universities of Alberta,
Calgary, Lethbridge and Manitoba, and by C3.ca, the Netera Alliance, CANARIE,
the Alberta Science and Research Authority, and the CFI.

This work was supported in part by grants 
from NSF: PHY-0244699 to Louisiana State University, 
and NSF PHY-0099568, NSF PHY-0244906 to Caltech. Additional funding
came from Caltech's Richard Chase Tolman Fund, 
NSERC and a Research Innovation Award of Research Corporation to LSU.
L.L. is an Alfred P. Sloan Fellow.

\end{document}